\documentstyle[preprint,aps,prl,amstex,psfig]{revtex}
\tightenlines

\DeclareMathSymbol{\square}       {\mathord}{AMSa}{"03}
\DeclareMathSymbol{\blacksquare}  {\mathord}{AMSa}{"04}
\DeclareMathSymbol{\lozenge}      {\mathord}{AMSa}{"06}
\DeclareMathSymbol{\blacklozenge} {\mathord}{AMSa}{"07}
\DeclareMathSymbol{\vartriangleright}{\mathrel}{AMSa}{"42}
\DeclareMathSymbol{\vartriangleleft} {\mathrel}{AMSa}{"43}
\DeclareMathSymbol{\blacktriangledown}  {\mathord}{AMSa}{"48}
\DeclareMathSymbol{\blacktriangleright} {\mathrel}{AMSa}{"49}
\DeclareMathSymbol{\blacktriangleleft}  {\mathrel}{AMSa}{"4A}
\DeclareMathSymbol{\vartriangle}        {\mathrel}{AMSa}{"4D}
\DeclareMathSymbol{\blacktriangle}      {\mathord}{AMSa}{"4E}
\DeclareMathSymbol{\triangledown}       {\mathord}{AMSa}{"4F}

\begin{document}

\draft
\title{Subharmonic bifurcation cascade of pattern oscillations caused by 
winding number increasing entrainment}
\author{Ch.~Jung, B.~Huke, and M.~L\"{u}cke}
\address{Institut f\"{u}r Theoretische Physik, Universit\"{a}t des 
Saarlandes, Postfach 151150, D-66041 Saarbr\"{u}cken, Germany}
\maketitle


\begin{abstract}
Convection structures in binary fluid mixtures are investigated for positive
Soret coupling in the driving regime where solutal and thermal
contributions to the buoyancy forces compete. Bifurcation properties of 
stable and unstable stationary square, roll, and crossroll (CR) structures and
the oscillatory competition between rolls and squares are determined 
numerically as a function of fluid parameters.
A novel type of subharmonic bifurcation cascade (SC) where the oscillation 
period grows in 
integer steps as $n\frac{2\pi}{\omega}$ is found and elucidated to be an 
entrainment process. 
\end{abstract}


%
\narrowtext
Many nonlinear dissipative systems that are driven away from 
equilibrium  by a stationary external forcing respond by spontaneous 
macroscopic  selforganization in time periodic oscillating 
patterns, e.g., in the form  of propagating waves but also in 
standing 
structures \cite{CH}. A system that,  depending on 
parameters, shows both 
forms is convection in binary fluid  mixtures confined 
between 
horizontal plates that impose a vertical 
temperature gradient: This forcing generates for negative 
thermodiffusive Soret coupling $\psi$ between temperature and 
concentration fields propagating convection waves 
\cite{CH}. 
On the other hand, for $\psi > 0$ periodic pulsations 
\cite{GPC85,MS91,DAC95,theory} of a 
standing square pattern have been seen such that the intensity 
oscillations are global and homogeneous along the lines of 
the squares. 
 
Here we report that and we elucidate how 
and why such pattern oscillations undergo a SC in which the 
period $\tau$ does not double but increases as  
$n\frac{2\pi}{\omega}$
with $n=1, 2, 3, 4, ...$. This peculiar dynamical 
behavior reflects a novel bifurcation pathway of global pattern oscillations 
that has not been reported before. To explain it 
we have to keep track of the bifurcation properties of three different 
stationary convective structures (squares, rolls, and 
CRs as explained below). These three states 
"organize" the dynamical behavior of the system, i.e., of the 
oscillatory convection. They allow to understand the SC as an 
entrainment process and to identify the integer $n$ in the SC as the 
number of windings around a stable CR state of increasing attraction 
which at the end of the SC becomes sufficiently strong to quench 
the oscillations. 
 
We have solved the full 3D Oberbeck-Boussinesq 
hydrodynamic field equations 
for convection in binary fluids with two different numerical 
methods: a finite-difference code \cite{HW65} 
augmented by a fast multigrid pressure iteration scheme 
\cite{CJ97} and 
a Galerkin expansion which also yields unstable 3D solutions. Top and 
bottom boundaries were rigid, impermeable, perfectly heat 
conducting plates. 
Using a $2 \times 2$ integration domain in $x, y$ with 
periodic boundary  
conditions the wavelength $\lambda=2$ \cite{dim} was fixed to 
the experimentally observed one \cite{GPC85,MS91,DAC95}. A few 
simulations were done for $16 \times 16$ systems with rigid, 
impermeable, 
heat insulating side walls to check the robustness of the 
patterns. We now discuss our findings. 
 
When the thermal stress, i.~e., the Rayleigh number is 
increased the primary stable convective 
structure for positive Soret coupling, $\psi > 0$, has the 
form of stationary squares for a  
wide range of Lewis numbers $L$ and Prandtl numbers  
$\sigma$ as predicted by bifurcation theory \cite{ClK91}.  
Then, convection in the form of pure rolls is an unstable solution. 
Close to the convective threshold the square structures can 
roughly be viewed to consist of a superposition of two orthogonal sets 
of straight parallel stationary convective rolls. We shall call 
them 
$x$-rolls with mode or flow intensity $X$ and wavevector 
oriented 
in $x$-direction and  $y$-rolls, respectively, with mode   
intensity $Y$ and wavevector in $y$-direction \cite{XY}.
In square convection the intensities of the two constitutive 
rolls are 
the same, $X_S = Y_S$. For the unstably coexisting pure 
$x$-roll 
solution one has $X_R > 0$ and $Y_R = 0$. The symmetry 
degenerate pure 
$y$-roll solution has $X_R = 0$ and $Y_R > 0$. 
 
We found that the stationary solution branches for 
squares and rolls are connected by branches of stationary 
CRs 
(lines with triangles in Fig.~\ref{bif}) that bifurcate out of the 
square  
branch and merge with increasing $r$ with the roll 
bifurcation 
branch. Bifurcation properties of these states were unknown 
so far despite 
experimental hints for their existence \cite{GPC85,BCC90}. 
In the CR state the two sets of $x$-rolls and $y$-rolls 
superimpose with different intensities, e.~g., $X_{CR} > 
Y_{CR} > 0$ but  
equal wave number $k_x = k_y$. These structures differ 
from the rectangular ($k_x \neq k_y$) CR patterns that 
grow in 
pure fluid convection at large $r$ out of rolls \cite{CB94}. 
The transition from CRs to pure rolls is marked by the 
vanishing of $Y_{CR}$. In Fig.~\ref{bif} we identify 
$X_{CR}$ by upwards  
($Y_{CR}$ by downwards) pointing triangles. The CR 
state with interchanged intensities and dominating $y$-rolls is 
symmetry degenerate with the former and corresponds to a
convective structure that is rotated by $90^{\circ}$.
In addition to the obvious magnitude relations of mode intensities 
in Fig.~\ref{bif} like, e.g., 
$0 \leq Y_{CR} \leq Y_S = X_S \leq X_{CR}  \leq X_R$
 we found for each $r$ shown there that 
$2 X_S \lesssim X_{CR} + Y_{CR} \lesssim X_R$,
$N_S \lesssim N_{CR} \lesssim N_R$,
and $w_R \leq w_{CR} \leq w_S \lesssim \sqrt{2} w_R$ \cite{smallr}. Here the
Nusselt number $N$ behaves as $N -1 \propto (X + Y)$ while the vertical
flow amplitude $w \propto (\sqrt{X} + \sqrt{Y})$. In the above relations
the CR quantities enter only where they 
exist, e.g.,  $N_{CR}$ starts out of the square state with $N_{CR}=N_S$,
grows with increasing $r$, and ends with $N_{CR}=N_R$ in the roll state.

For not too large Lewis numbers, i.e., for {\it liquid} 
mixtures the 
competition between square and pure roll convection leads to 
the 
appearance of pulsations of $X(t)$ and $Y(t)$ 
\cite{GPC85,MS91,DAC95}. These  
oscillations grow in a supercritical Hopf bifurcation 
out of the square state thereby rendering the latter unstable. 
Fig.~\ref{os2} shows  
for several Rayleigh numbers 
that $X(t)$ and $Y(t)$ oscillate in opposite phase around 
a common mean value given by the unstable square state 
($\Box$  
in Fig.~\ref{os2}) such that always $X(t) + Y(t) \cong 2 X_S$. 
The $x$-roll intensity $X(t)$ of the pulsating pattern grows 
or decreases on 
cost of the $y$-roll intensity $Y(t)$, however, without ever 
going to zero. Thus 
the two roll sets never die out completely or reverse their
turning  
direction during the oscillations. 

In Fig.~\ref{int} we show grey-scale snapshots of the topview 
shadowgraph intensity distribution $I(x,y)$ taken at times 
marked in Fig.~\ref{os2} by A ($X$ minimal; $y$-rolls 
dominant), B ($X = Y$; 
squares), and C ($X$ maximal; $x$-rolls dominant). After 
half an oscillating period the pulsating convective structure 
appears to be rotated by  
$90^{\circ}$ in the $x-y$ plane since   
${\cal F}(x,y,z;t+\frac{\tau}{2}) = {\cal F}(y,-x,z;t)$ 
holds for the fields 
${\cal F}=\delta T$ and $\delta C$ of temperature and 
concentration,  
respectively. Furthermore they show always 
the mirror glide symmetry  
${\cal F}(x,y,z;t) = -{\cal F}(x+ 
\frac{\lambda}{2},y+\frac{\lambda}{2},1-z;t)$ 
\cite{symmetry}. 
The intensity distribution $I(x,y)$ was calculated with the 
formula \cite{INT} 
\begin{equation} 
\label{intensity} 
\hspace*{-1cm}I = I^T + I^C \sim  r\left( \partial^2_x + 
\partial^2_y \right) \, 
\int_0^1  dz \,(\delta T + b \, \delta C). 
\end{equation} 
The weighting factor $b$ = - 0.43 is for the parameter 
combination 
$\psi = 0.23, \, L = 0.0045, \, \sigma = 27$ that can be 
realized by 
ethanol-water mixtures \cite{MS91}. The shadowgraph intensity profile 
along the  
dotted line in Fig.~\ref{int} B is shown in the bottom 
right corner together with the contributions $I^T$ and $I^C$ 
to it from the 
temperature field and the concentration field, respectively. 
Narrow  
concentration plumes caused by the smallness of $L$ 
generate a 
narrow central bright stripe in B at the upflow location $x 
\cong 0.5$ and a peak in $I$ on top of the smooth contribution from 
$\delta T$. This  
characteristic structure should be easily 
accessible in experiments.  

Close to the Hopf bifurcation the oscillations are harmonic 
and of  
small amplitude (Fig.~\ref{os2}a). With increasing thermal 
driving $r$ the  
frequency decreases roughly linearly \cite{MS91} (for 
example, in  
Fig.~\ref{bif} the Hopf frequency is $\omega_H=0.251$ and 
$\omega(r=1.173)=0.0586$). Furthermore, and more 
importantly, 
with increasing amplitude the oscillation becomes more 
anharmonic 
and relaxational in Fig.~\ref{os2}b, c; see also 
Ref.~\cite{GPC85,MS91}: While the  
system rapidly sweeps through the square state ($\Box$) it 
spends more  
and more time in the vicinity of the roll state ($\circ$). 
The change from harmonic in Fig.~\ref{os2}a to  
strongly relaxational oscillations in Fig.~\ref{os2}c is also 
documented in the right  
column of Fig.~\ref{os2}. There we show phase space plots 
$X, Y$  
versus $\dot {X}, \dot {Y}$ associated with the time 
histories of  
$X(t), Y(t)$ in the left column. 
 
At larger $r$ the system gets attracted into one of the   
CR fixed points that have become stable shortly below 
the 
$r$-interval marked SC in Fig.~\ref{bif}. In this interval we 
have  
observed a novel subharmonic bifurcation cascade in which 
the CR 
attractors entrain the oscillations: First the CR attractors 
deform the phase trajectory (Fig.~\ref{os2}c). Then with 
increasing $r$ the oscillations execute an  
increasing number of windings  
around the CR states (triangles). In Figs.~\ref{os2}c - f 
the  
winding number  
around a CR fixed point increases from $n$ = 1 to $n$ 
= 4 and the  
period $\tau$ of the oscillations increases from  
$\frac{2\pi}{\omega}$ to $n \frac{2\pi}{\omega}$ in integer 
steps. 
This increase of the winding number in the SC continues 
beyond $n=4$; we  
have found also $n = 5$. However, the control parameter 
interval  $\delta r_n$  
for an $n$-cycle becomes so narrow --- $\delta r_n \cong 
(1.6, 0.6, 0.2) 10^{-4}$ 
for $n=(2, 3, 4)$ --- that our numerical resources were not  
sufficient to resolve the SC further.  
But we think that the SC is a robust, experimentally 
accessible phenomenon;  
Ref.~\cite{GPC85} contains a hint for a 2-cycle. Beyond the 
SC  
interval in  Fig.~\ref{bif} the system gets   
attracted into one  
of the CR fixed points. The transition  between the 
oscillations   
and the CR state is slightly hysteretic. Upon reducing 
$r$ the  
system remains in the CR  
state until below the SC interval and then a  
transition to an $n = 1$ oscillation occurs. So in a small $r$-
interval there is bistable coexistence of oscillations and  
stationary CRs. 
 
In Fig.~\ref{lps} we present phase diagrams in 2D planes of 
the $r - \psi - L - \sigma$ parameter 
space. Consider first the $L-r$ plane 
of Fig.~\ref{lps}a. Upon crossing the bifurcation threshold of 
the conductive 
state squares (rolls) become stable at $L \lesssim 0.45$ ($L 
\gtrsim 0.45$). 
At smaller $\psi$ we could compare this stability boundary 
between squares 
and rolls with results from Ref.  \cite{ClK91} and found 
excellent agreement. The oscillatory regime extends only 
up to $L\approx 0.02$.  This, by the way,  
explains why oscillations could not be found in the 
experiments of  
Ref.~\cite{BCC90}. For $L>$ 0.02 squares transfer their 
stability 
directly to CRs when increasing $r$.   

Consider now the $\psi-r$ plane of Fig.~\ref{lps}b where 
stable 3D structures  
are predicted to appear at onset already for  
$\psi \gtrsim 10^{-7}$ \cite{ClK91}. For the smallest $\psi$ 
that we 
have analyzed, $\psi=0.03$, we found already the full 
sequence of squares, 
oscillations, and CRs with increasing $r$. With growing 
Soret coupling $\psi$ the stable existence range of each of 
these 3D structures 
that are peculiar to mixtures widens on cost of the 2D roll 
states. 
The onset of the latter is shifted upward to larger $r$: With  
increasing $\psi$ it requires higher thermal stress to generate 
the 
flow intensity for which advective mixing is strong enough to 
eliminate effectively the Soret induced concentration 
variations that  
ultimately cause and/or stabilize the 3D structures. 
Convection in well  
mixed binary fluids at large $r$ is not much different from 
one-component 
fluids. Furthermore, our observation that the ratio of mean squared 
convective concentration and temperature variations that determine 
solutal and thermal buoyancy contributions, respectively, 
is large (small) compared to one at small (large) $r$
sheds light on the oscillatory and/or stationary pattern 
competition at intermediate $r$: At small (large) $r$ squares 
(rolls) are stably sustained by solutally (thermally) dominated buoyancy 
forces and in the crossover driving regime in between where neither 
squares nor rolls are stable there is oscillatory and/or stationary 
pattern competition. As an aside we mention that the advective concentration
(heat) transport and the convective concentration (temperature) variation is
slightly larger (smaller) in squares than in rolls.
  
Studying the Prandtl number dependence in  Fig.~\ref{lps}c 
we found that  
oscillations disappear at $\sigma \lesssim 0.2$ and that rolls 
appear already 
slightly above $r=1$. Increasing $\sigma$ the 
existence range of 
squares, oscillations, and CRs expands to higher $r$. 
We have also investigated parameters $L, \sigma = {\cal 
O}(1)$ which 
apply to {\it gas} mixtures \cite{LA97}. Here typically 
rolls are stable slightly above onset. Knowledge \cite{IP98} of  
their stability domain is useful in relation to the competing  
spiral defect chaos that has recently been observed also in 
gas mixtures \cite{LA97}. 


{\it Acknowledgments\/} --- This work was supported by the 
Deutsche Forschungsgemeinschaft. The 
H\"{o}chstleistungsrechenzentrum J\"{u}lich  
provided computing time. Numerical support by M.~Kamps and
M.~F\"{u}cker is gratefully acknowleged.


 
 
\begin{figure}
\caption[] 
{Bifurcation diagram of x-roll intensity $X$ and $y$-roll 
intensity $Y$  
versus $r$ for squares (squares), rolls (circles), and  
crossrolls (triangles). Full (dashed) lines with filled (open) 
symbols denote  
stable (unstable) states. Full lines delimiting the hatched area 
mark maxima  
and minima of oscillations of $X(t)$ and $Y(t)$. 
A subharmonic bifurcation cascade occurs in the $r$-interval  
marked SC (width expanded for better visibility).  
For the parameters \cite{MS91} $\psi= 0.23, L = 0.0045, \sigma = 27$ used 
here stable  
squares and unstable rolls bifurcate with wave number $k = \pi$ 
out of the  
conductive state at $r = 0.0124$ while the critical Rayleigh 
number is  
$r_c = \frac{720}{R_c^0}\,\frac{L}{\psi} = 0.00825$.}   
\label{bif}    
\end{figure} 
 
 
\begin{figure}
\caption[] 
{Evolution of the oscillatory dynamics with increasing $r$  
from top to bottom. Parameters as in Fig.~\ref{bif}. 
(a) and (b) are states located shortly above the Hopf threshold 
and right 
below the CR bifurcation, respectively. (c) - (f) are SC 
states. 
The right column contains phase space plots of $X, Y$ versus  
$\dot {X}, \dot {Y}$ for
$X(t), Y(t)$ shown in the left column with arbitrary common 
units versus 
reduced time $n \frac {t}{\tau}$. In the SC states $n$ 
is the number  
of windings around each of the stable CR fixed points 
(triangles). 
Squares and circles indicate unstable square and roll states, 
respectively.} 
\label{os2}    
\end{figure} 
 
 
\begin{figure}
\caption[] 
{Grey-scale snapshots of topview shadowgraph intensity 
distribution 
$I(x, y)$ evaluated with Eq.~(\ref{intensity}) for the oscillatory 
state 
of Fig.~\ref{os2}b at times marked by A ($y$-rolls dominant), 
B (squares), 
and C ( $x$-rolls dominant). To elucidate the contributions 
from temperature 
and concentration we show in the bottom right corner the 
profiles of 
$I^T, I^C$, and $I=I^T + I^C$ along the dotted line in B. 
Maximal 
up (down) flow is located in B at $x=y\simeq$ 0.5 (1.5).} 
\label{int}    
\end{figure} 
 
 
\begin{figure}
\caption[] 
{Phase diagrams of stable convective states. The narrow SC domain 
is not shown. Parameters are $\psi=0.23$ (a, c),  
$L=0.0045$ (b, c), and $\sigma=27$ (a,b). Fig.~\ref{bif} 
shows for them 
a bifurcation diagram along the vertical dotted lines in (a, b). 
The ordinate scale is logarithmic for $r<1$ to show  
the bifurcation threshold $r_{stat}(k=\pi; \psi, L)$ of the 
conductive state (cond) which was determined with a shooting method. 
Phase boundaries in (a) between crossrolls and rolls from finite 
difference (full line) and Galerkin (dashed line) calculations are 
quantitatively unreliable  where they disagree at small $L$ 
combined with large $r$, i. e., when narrow concentration 
boundary layers occur in the fluid.\label{lps}}   
\end{figure} 


\begin{thebibliography}{99}
\bibitem{dim}
We scale lengths by the thickness $d$ of the fluid layer and times
by the vertical heat diffusion time $d^2/\kappa$. Thermal
diffusivity $\kappa$, kinematic viscosity $\nu$, and concentration
diffusion coefficient $D$ define Prandtl number $\sigma = \nu/\kappa$
and Lewis number $L = D/\kappa$. Furthermore, we use the reduced
Rayleigh number $r = R/R^0_c$ where $R^0_c$ is the critical
one for onset of convection in a pure fluid. The finite-difference 
calculations were done with homogeneous spatial resolution of 0.05. The
Galerkin expansions included wave numbers up to $16\pi$ in each direction.
\bibitem{XY} 
In Figs.~\ref{bif} and \ref{os2} we present $X=|I_{1 0}|^2$ and 
$Y=|I_{0 1}|^2$ where $I_{n m}$ are Fourier modes of the topview
shadowgraph intensity $I(x,y)$ defined in Eq.~(\ref{intensity}). 
Taking instead (Fourier mode) intensities of the velocity fields in
$x$- and $y$-direction yields practically identical graphs.
\bibitem{smallr}
At small $r$ close to threshold we found  
$X_R \lesssim 2 X_S$, $N_R \lesssim N_S$, and $\sqrt{2} w_R \lesssim w_S$
in accordance with a simple amplitude equation that is constructed such as 
to yield linear growth for
$X(r)$ and stable squares \cite{CH}. However with increasing $r$ the 
deviations grow: the intensities of both squares and rolls curve upwards 
but in different ways such that at $r \gtrsim 0.6$ the above magnitude
relations reverse without however rolls becoming stable.
\end{thebibliography}
\end{document}